\documentclass[prd,twocolumn,superscriptaddress]{revtex4}
\usepackage{amsmath}
\usepackage{adjustbox}
\usepackage{gensymb}
\usepackage{amsfonts}
\usepackage{CJKutf8}
\usepackage{graphicx}
\usepackage{hyperref}
\usepackage{amssymb}
\usepackage{cancel}
\usepackage{color}
\usepackage{amscd}
\usepackage{bm}
\usepackage{adjustbox}
\usepackage{mathrsfs}

\makeatletter
\renewcommand{\maketag@@@}[1]{\hbox{\m@th\normalsize\normalfont#1}}%
\makeatletter
\begin{document}
\begin{CJK*}{UTF8}{gbsn}
\title{Thermodynamic Volumes of Extended Black Hole Thermodynamics with Auxiliary Gauge Fields}
\author{Zili Mou}
\email{202421101027@bnu.edu.cn}
\affiliation{School of Physics and Astronomy, Beijing Normal University, Beijing 100875, China}

\author{Yu Tian}
\email{ytian@ucas.ac.cn}
\affiliation{School of Physical Sciences, University of Chinese Academy of Sciences, Beijing 100049, China}
\affiliation{Institute of Theoretical Physics, Chinese Academy of Sciences, Beijing 100190, China}

\author{Hongbao Zhang}
\email{hongbaozhang@bnu.edu.cn}
\affiliation{School of Physics and Astronomy, Beijing Normal University, Beijing 100875, China}
\affiliation{Key Laboratory of Multiscale Spin Physics, Ministry of Education, Beijing Normal University, Beijing 100875, China}

\begin{abstract}

 By applying the Iyer-Wald formalism to the modified Lagrangian form for the gravitational system, where the coupling constants appearing in the original Lagrangian form, such as the cosmological constant, are promoted to the dynamical scalar fields by introducing the corresponding auxiliary gauge fields, we derive the first law of the extended black hole thermodynamics, where the resulting thermodynamic volumes as the conjugate thermodynamic variables to the coupling constants are nevertheless independent of the auxiliary gauge fields and exactly the same as those obtained before via varying the coupling constants by brute force. In addition, it is further shown that such volumes can also be expressed in terms of the auxiliary gauge fields and evaluated on the black hole horizon, validating the proposal made in the previous literature.  
\end{abstract}

\maketitle

\section{Introduction}
Compared to the ordinary black hole thermodynamics, the extended black hole thermodynamics allows the coupling constants $\alpha^i$ such as the cosmological constant to vary with the conjugate thermodynamic variables (usually dubbed as thermodynamic volumes) given by $\mathcal{V}_i=(\frac{\partial M}{\partial \alpha^i})_{S,J,\cdots}$\cite{Kastor:2009wy,Dolan:2010ha,Kubiznak:2012wp,Dutta:2022wbh,Kastor:2010gq,Sinamuli:2017rhp,Xiao:2023lap}. Such an extension has led to wide-ranging applications in many research directions \cite{Kubiznak:2016qmn,Mann:2025xrb,Wei:2015iwa,Wei:2019uqg,Cong:2021fnf,Ahmed:2023snm,Zhang:2025dgm,Caceres:2016xjz,Couch:2016exn,AlBalushi:2020rqe,Harlow:2022ich,Gwak:2017kkt}. But nevertheless, the extended black hole thermodynamics had faced two longstanding issues. For one thing, the aforementioned coupling constants are constants that define the theory and are conceptually fixed. Thus there is no natural physical interpretation for their variation. For another, in order to obtain the thermodynamic volumes through the above thermodynamic relation, one is required to first calculate out the explicit expression for the black hole mass. A direct expression for these thermodynamic volumes is lacking.

With regard to the first issue, the authors in \cite{Hajian:2023bhq}, among others, constructed a general theoretical framework, where all coupling constants are successfully promoted to be dynamical scalar fields $\alpha^i(x)$ by introducing an auxiliary Abelian gauge field $\mathbf{A}_i$ for each constant. The resulting theory is on-shell equivalent to the original theory because each $\alpha^i(x)$ that satisfying the equation of motion turns out to be a constant over the whole spacetime. 

 Furthermore, inspired by the analogy with the electric potential, the authors in \cite{Hajian:2023bhq} also tried to resolve the above second issue by proposing a direct formula for the thermodynamic volumes as 

\begin{equation}
   \mathcal{V}_i=-\int_\mathcal{B}\xi\cdot \mathbf{A}_i,
    \label{chemi}
\end{equation}
where $\xi$ denotes the Killing vector field normal to the black hole horizon, and $\mathcal{B}$ denotes the bifurcation surface of the black hole horizon\footnote{Note that $\xi=0$ at the bifurcation surface, so $\boldsymbol{A}_i$ is generically non-smooth over there. With this in mind, we had better interpret the above expression as a limit from outside the black hole.}.
However, such a proposal was only tested by one example therein but not well justified on general grounds.

On the other hand, a direct expression for the thermodynamic volumes has most recently been achieved in  \cite{Xiao:2025gle} as the sum of two terms
\begin{align}
    \mathcal{V}_i = \mathcal{V}_i^{(1)}+\mathcal{V}_i^{(2)},
    \label{v1v2}
\end{align}
where the first term is a bulk term, while the second term is a boundary term at infinity. However, the whole derivation is not formulated within a theoretical framework that allows for a natural variation of the coupling constants as done in \cite{Hajian:2023bhq}.

The purpose of this paper is two-fold. First, by working with the extended black hole thermodynamics with the auxiliary gauge fields, we derive the equivalent expression for the thermodynamic volumes to Eq. (\ref{v1v2}) obtained in \cite{Xiao:2025gle}. Second, we show that such an expression can also be formulated in terms of Eq. (\ref{chemi}), justifying the proposal made in \cite{Hajian:2023bhq}.

The structure of this paper is organized as follows. In Section \ref{sec1}, we will briefly review the derivation of the first law of the ordinary black hole thermodynamics based on the background subtraction within the Iyer-Wald formalism. Subsequently, we present the derivation of the extended black hole thermodynamics within the theoretical framework with the coupling constants promoted to the dynamical scalar fields by the introduction of the auxiliary gauge fields in Section \ref{sec:III}, where with the resulting auxiliary gauge field independent expression for the volume terms, we further derive its equivalent expression in terms of the auxiliary field by an appropriate gauge choice. We conclude our paper in the last section.

\section{Iyer-Wald formalism, background subtraction and the first law of ordinary black hole thermodynamics } \label{sec1}

In this section, we shall provide a concise review of the derivation of the first law of ordinary black hole thermodynamics based on the Iyer-Wald formalism \cite{Wald:1993nt,Iyer:1994ys,Iyer:1995kg} supplemented with the background subtraction procedure.

In what follows, for definiteness we focus on a diffeomorphism invariant Lagrangian form $\mathbf{L}$ for pure gravity. We have its variation in the following form
\begin{equation}
    \delta\mathbf{L} =\boldsymbol{\epsilon} E^{ab}\,\delta g_{ab} + d\boldsymbol{\Theta},
    \label{gelvariform}
\end{equation}
where $\boldsymbol{\epsilon}$ is the volume element, $ {E}^{ab}=0$ is the equation of motion,  and $\bm{\Theta}$ is called the pre-symplectic potential.  The standard Noether current associated with a vector field $\xi$ is defined as
\begin{equation}
    \mathbf{J}_\xi \equiv X_\xi\cdot\boldsymbol{\Theta} - \xi\cdot\mathbf{L},
    \label{gelj}
\end{equation}
where $X_\xi\equiv\int d^dx\sqrt{-g}\mathcal{L}_\xi g_{ab}\frac{\delta}{\delta g_{ab}}$ and $\bm{\Theta}$
 are understood respectively as a vector field and a one-form on the configuration space of our dynamical fields with the dot denoting the contraction with the first index of the involved form. It is easy to show
 \begin{equation}
     d\mathbf{J}_\xi=-\boldsymbol{\epsilon}E^{ab}\mathcal{L}_\xi g_{ab}.
     \label{conserve}
 \end{equation}
Then one can further obtain
\begin{equation}
    d(\mathbf{J}_\xi-\mathbf{C}_\xi)=\xi^a\mathcal{B}_a,
\end{equation}
where $\mathbf{C}_\xi=c_\xi\cdot\boldsymbol{\epsilon}$ with ${c_\xi}^a=-2\xi^c{E}^a_{\;c}$ is the constraint which vanishes on shell, and   $\mathcal{B}_a=2\boldsymbol{\epsilon}\nabla_c{E}^c_{\;a}$ vanishes automatically as the Bianchi identity due to the fact that $\xi$ can be chosen in an arbitrary manner. Accordingly, the Noether current can be expressed as
\begin{equation}
    \mathbf{J}_\xi =\mathbf{C}_\xi+ d\mathbf{Q}_\xi
\end{equation}
with $\mathbf{Q}_\xi$ called the Noether charge.

With the bulk pre-symplectic form defined as $\boldsymbol{\omega}\equiv\delta\boldsymbol{\Theta}$ and the definition $\mathbf{k}_\xi \equiv\delta\mathbf{Q}_\xi-\xi\cdot\boldsymbol{\Theta}$, one can derive the fundamental identity
\begin{align}
    d \mathbf{k}_\xi =  -X_\xi\cdot\boldsymbol{\omega}-\xi\cdot\boldsymbol{\epsilon}E^{ab}\delta g_{ab}-\delta\mathbf{C}_\xi.
    \label{fundamental}
\end{align}
If we evaluate Eq. \eqref{fundamental} on top of the solution space with $X_\xi=0$ and restrict those variations which satisfy the linearized equations of motion, then all the three terms on the right-hand side of \eqref{fundamental} vanish. As a result, we have
\begin{align}
    d \mathbf{k}_\xi =  0.
    \label{fundamental2}
\end{align}
In particular, for a given stationary black hole, one can choose $\xi$ as the properly normalized Killing vector field normal to the black hole horizon as $\xi=\frac{\partial}{\partial t}+\Omega_H \frac{\partial}{\partial \phi}$ with $\Omega_H$ the angular velocity at the horizon. Here we work implicitly with the four-dimensional spacetime for definiteness. The generalization to other spacetime dimensions is supposed to be straightforward unless otherwise specified. Now integrating Eq. (\ref{fundamental2}) over a spatial hypersurface $\Sigma$ that emanates from the bifurcation surface $\mathcal{B}$ at which $\xi=0$ and ends on the intersection surface $\mathcal{C}$ with a timelike hypersurface $\Gamma$ which will be taken to the infinity at the end of the day, we have
\begin{equation}
   \int_{\mathcal{C}}\mathbf{k}_\xi=\int_{\mathcal{B}}\mathbf{k}_\xi=T\delta S,
    \label{id1}
\end{equation}
where $T=\frac{\kappa}{2\pi}$ and $S=\frac{2\pi}{\kappa}\int_\mathcal{B}\mathbf{Q}_\xi$ represent the black hole temperature and entropy, respectively. 
To obtain the first law of black hole thermodynamics
\begin{equation}
  \delta [H_\xi]= T\delta S
    \label{1stlaw}
\end{equation}
with $[H_\xi]$ the finite Arnowitt-Deser-Misner (ADM) Hamiltonian conjugate to $\xi$, we like to follow the strategy developed in \cite{Guo:2025ohn} to perform a background subtraction procedure with respect to a reference spacetime. As such, we first massage $\mathbf{\Theta}$ on $\Gamma$ into the following form
\begin{equation}
    \mathbf{\Theta}\big|_\Gamma = -\delta\mathbf{B} + d\mathbf{C} + \mathbf{F},\label{sec2theta1}
\end{equation}
whereby we have
\begin{eqnarray}
    k_\xi&=&\delta \mathbf{Q}_\xi+\delta(\xi\cdot\mathbf{B})-\mathcal{L}_\xi\mathbf{C}+d(\xi\cdot\mathbf{C})-\xi\cdot\mathbf{F}\nonumber\\
    &=&\delta \mathbf{Q}_\xi+\delta(\xi\cdot\mathbf{B})-\mathcal{L}_{X_\xi}\mathbf{C}+d(\xi\cdot\mathbf{C})-\xi\cdot\mathbf{F}\nonumber\\
    &=&\delta (\mathbf{Q}_\xi+\xi\cdot\mathbf{B}-{X_\xi}\cdot\mathbf{C})+d(\xi\cdot\mathbf{C})-\xi\cdot\mathbf{F},
\end{eqnarray}
where we have used the diffeomorphism covariance of $\mathbf{C}$, namely $\mathcal{L}_\xi\mathbf{C}=\mathcal{L}_{X_\xi}\mathbf{C}$ in the second line and the Cartan identity with $X_\xi=0$ in the last line. Now denote $H_\xi=\int_\mathcal{\infty}\mathbf{Q}_\xi+\xi\cdot\mathbf{B}-{X_\xi}\cdot\mathbf{C}$, then Eq. (\ref{id1}) reduces to 
\begin{equation}
    \delta H_\xi-\int_\mathcal{\infty}\xi\cdot\mathbf{F}=T\delta S.
\end{equation}
On the other hand, for the reference spacetime, where there is no inner boundary for the spatial hypersurface, we have 
\begin{equation}
\delta H^0_\xi-\int_\mathcal{\infty}\xi\cdot\mathbf{F}^0=0
\end{equation}
with the superscript $0$ representing the corresponding quantities in the reference spacetime. With $[H_\xi]\equiv H_\xi-H_\xi^0$, we end up with the first law of thermodynamics (\ref{1stlaw}), because it is shown in \cite{Guo:2025ohn} that not only is $[\int_\mathcal{\infty}\xi\cdot \mathbf{F}]=0$  satisfied for both asymptotically flat spacetimes and asymptotically AdS spacetimes, but also the resulting $[H_\xi]$ is finite.

\section{Auxiliary gauge fields, the first law of the extended black hole thermodynamics, and volume terms}\label{sec:III}
For our purpose, let us specify the following Lagrangian form
\begin{equation}
    \mathbf{L} =\mathbf{L}_0+ \alpha _i\mathbf{L}_i,
    \label{lagrangian}
\end{equation}
where $\mathbf{L}_0$ represents the Einstein-Hilbert action, and $\alpha ^i$ denote the cosmological constant and the effective Wilsonian coefficients from the high derivative corrections. It turns out that the Smarr formula for such a theory also involves $\alpha^i$, which motivates one to treat $\alpha^i$ also as thermodynamic variables, leading to the extended black hole thermodynamics.  To provide a natural mathematical framework in which $\alpha^i$ can be legitimately varied, the authors in \cite{Hajian:2023bhq} instead work with the following modified Lagrangian form
\begin{equation}
    \mathbf{L}^\prime =\mathbf{L}_0+\alpha ^i(x)\bigg(\mathbf{L}_i-d\mathbf{A}_i\bigg),
    \label{lagrangian2}
\end{equation}
where each $\alpha^i$ in the original theory is promoted to a dynamical scalar field $\alpha^i(x)$ and an auxiliary gauge field $\mathbf{A}_i$ is introduced for each $\alpha^i(x)$. Below we shall adapt the Iyer-Wald formalism to such a modified Lagrangian form such as to derive the first law of the extended black hole thermodynamics as well as the explicit expression for the thermodynamic variables conjugate to $\alpha^i$. 

With $\delta \mathbf{L}_0=\boldsymbol{\epsilon}{E_0}^{ab}\delta g_{ab}+d\mathbf{\Theta}_0$ and $\delta \mathbf{L}_i=\boldsymbol{\epsilon}{E_i}^{ab}\delta g_{ab}+d\mathbf{\Theta}_i$, we have 
\begin{align}
       &\delta\mathbf{L}^\prime =\boldsymbol{\epsilon} E^{\prime ab}\,\delta g_{ab}+\mathbf{E^\prime}^i\wedge\delta\mathbf{A}_i+\mathbf{E^\prime}_i\delta\alpha^i+ d\boldsymbol{\Theta}^\prime,
       \label{gelvariform2}
\end{align}
where
\begin{eqnarray}
    &E^{\prime ab}&=E^{ab}- D^{ab}\nonumber,\\
       &\mathbf{E^\prime}^i&=d\alpha^i\nonumber,\\
       &\mathbf{E^\prime}_i&=\mathbf{L}_i-d\mathbf{A}_i\nonumber,\\
       &\boldsymbol{\Theta}^\prime&=\boldsymbol{\Theta}- \alpha^i\delta \mathbf{A}_i-\mathbf{\Psi}.
       \label{eom}
\end{eqnarray}
Here $E^{ab}$ and $\mathbf{\Theta}$ correspond to the counterparts arising from the original Lagrangian form with $\alpha^i$ fixed, namely $E^{ab}\equiv {E_0}^{ab}+\alpha^i{E_i}^{ab}$, and $\boldsymbol{\Theta}\equiv\boldsymbol{\Theta}_0+\alpha^i\boldsymbol{\Theta}_i$. In addition, $D^{ab}$ and $\mathbf{\Psi}$ come from the decomposition of  $\mathbf{E}'^i\wedge\boldsymbol{\Theta}_i=\boldsymbol{\epsilon}D^{ab}\delta g_{ab}+d\mathbf{\Psi}$. Note that the solution to $\mathbf{E^\prime}^i=0$ is given obviously by the constant $\alpha^i$, which follows $D^{ab}=0$ and $d\mathbf{\Psi}=0$. On the other hand, we have $d\mathbf{L}_i=0$ automatically, which implies that not only is the solution to $\mathbf{E^\prime}_i=0$ existing, but also determined completely by the on-shell configuration $(g_{ab},\alpha^i)$ up to an arbitrary gauge transformation $\mathbf{A}_i\rightarrow \mathbf{A}_i+d\mathbf{\Lambda}_i$. So the resulting reduced on-shell configuration space $(g_{ab},\alpha_i)$ in the modified theory comprises the on-shell configuration spaces in the original theory with varied $\alpha_i$.

Now for the aforementioned gauge transformation, we can define the corresponding Noether current as follows
\begin{equation}
    \mathbf{J}'_\mathbf{\Lambda}=X_\mathbf{\Lambda}\cdot \mathbf{\Theta}',
\end{equation}
where $X_\mathbf{\Lambda}\equiv\int d^dx\sqrt{-g}d\mathbf{\Lambda}_i\frac{\delta}{\delta \mathbf{A}_{i}}$. Then we have
\begin{eqnarray}
     d\mathbf{J}'_\mathbf{\Lambda}&=&-\mathbf{E}'^i\wedge d\mathbf{\Lambda}_i
     =d(\mathbf{E}'^i \wedge \mathbf{\Lambda}_i)-d\mathbf{E}'^i\wedge \mathbf{\Lambda}_i.
\end{eqnarray}
The arbitrariness of our gauge transformation implies the corresponding Bianchi identity $d\mathbf{E}'^i=0$ as it should be the case. Accordingly, we have
\begin{equation}
    \mathbf{J}'_\mathbf{\Lambda}=\mathbf{E}'^i\wedge \mathbf{\Lambda}_i+d\mathbf{Q}_\mathbf{\Lambda}
\end{equation}
   with the Noether charge is given by 
   \begin{equation}
       \mathbf{Q}_\mathbf{\Lambda}=-\alpha^i \mathbf{\Lambda}_i.
   \end{equation}
Similarly, we can define the Noether current associated with the diffeomorphism for the modified theory as follows

\begin{equation}
    \mathbf{J}^\prime_\xi \equiv X'_\xi\cdot\boldsymbol{\Theta}^\prime - \xi\cdot\mathbf{L}^\prime,
    \label{gelj2}
\end{equation}
where $X'_\xi\equiv \int d^dx \sqrt{-g}\mathcal{L}_\xi\phi \frac{\delta}{\delta \phi}$ with $\phi$ collectively denoting $(g_{ab},\alpha^i,\mathbf{A}_i)$.
Whence we have
\begin{eqnarray}
   d\mathbf{J}^\prime_\xi&=& -\epsilon{E^\prime}^{ab}\mathcal{L}_\xi g_{ab}-\mathbf{E}'^i\wedge\mathcal{L}_\xi\mathbf{A}_i-\mathbf{E}'_i\mathcal{L}_\xi\alpha^i\nonumber\\
   &=&d\mathbf{C}^\prime_\xi+\xi^a\mathcal{B}^\prime_a,
\end{eqnarray}
where the constraint is given by $\mathbf{C}^\prime_\xi=c^\prime_\xi\cdot\boldsymbol{\epsilon}+\mathbf{E}'^i\wedge(\xi\cdot\mathbf{A}_i)$ with  ${c^\prime_\xi}^a=-2\xi^c{E^\prime}^a_{\;c}$ and $\mathcal{B}^\prime_a=2\boldsymbol{\epsilon}\nabla_c{E^\prime}^c_{\;a}-(E^i)_a\mathbf{L}_i=0$ by the Bianchi identity. As a result, we can write $\mathbf{J}^\prime_\xi$ as
\begin{equation}
    \mathbf{J}^\prime_\xi =\mathbf{C}^\prime_\xi+ d\mathbf{Q}^\prime_\xi.
\end{equation}
The fundamental identity for the modified theory is
\begin{eqnarray}
    d \mathbf{k}^\prime_\xi =&-&X'_\xi\cdot\boldsymbol{\omega}^\prime-\xi\cdot\boldsymbol{\epsilon}{E'}^{ab}\delta g_{ab}\nonumber\\
    &-&\xi\cdot(\mathbf{E^\prime}^i\wedge\delta\mathbf{A}_i)-\xi\cdot\mathbf{E^\prime}_i\delta\alpha^i-\delta\mathbf{C^\prime}_\xi.
    \label{funda2}
\end{eqnarray}
Here $\boldsymbol{\omega}^\prime\equiv\delta\boldsymbol{\Theta}^\prime=\boldsymbol{\omega}+ \delta\alpha^i\wedge\boldsymbol{\Theta}_i- \delta\alpha^i\wedge\delta \mathbf{A}_i-\delta\mathbf{\Psi}$ with $\boldsymbol{\omega}\equiv\delta\boldsymbol{\Theta}_0+\alpha^i\delta\boldsymbol{\Theta}_i$ for the original theory.

Next we like to express $d\mathbf{k}'_\xi$ in terms of $d\mathbf{k}_\xi$. To this end, note that 
\begin{eqnarray}
    \mathbf{J}^\prime_\xi&=&\mathbf{J}_\xi- \alpha^id(\xi\cdot \mathbf{A}_i)-X_\xi\cdot\mathbf{\Psi}\nonumber,\\
    \mathbf{C}^\prime_\xi&=&\mathbf{C}_\xi+2\xi^cD^a_{\;c}\boldsymbol{\epsilon}_{a\dots}+\mathbf{E}'^i\wedge(\xi\cdot\mathbf{A}_i).
    \label{JC}
\end{eqnarray}
Thus we have
\begin{eqnarray}
    &&d\mathbf{Q}^\prime_\xi-d\mathbf{Q}_\xi=\mathbf{J}^\prime_\xi-\mathbf{J}_\xi-(\mathbf{C}^\prime _\xi-\mathbf{C}_\xi)=\nonumber\\
   &&-2\xi^cD^a_{\;c}\boldsymbol{\epsilon}_{a\dots}-\alpha^id(\xi\cdot \mathbf{A}_i)-\mathbf{E}'^i \wedge (\xi\cdot\mathbf{A}_i)-X_\xi\cdot\mathbf{\Psi},\nonumber\\
   \label{charge}
\end{eqnarray}
which further gives rise to 
\begin{eqnarray}
    d\mathbf{k}^\prime_\xi-d\mathbf{k}_\xi =&&\delta (d\mathbf{Q}^\prime_\xi -d\mathbf{Q}_\xi)-d(\xi\cdot(\boldsymbol{\Theta}^\prime-\boldsymbol{\Theta}))\nonumber\\
    =&&-2\xi^c\delta(D^a_{\;c}\boldsymbol{\epsilon}_{a\dots})-\delta\alpha^i d(\xi\cdot\mathbf{A}_i)\nonumber\\
    &&-\delta\mathbf{E}'^i \wedge(\xi\cdot\mathbf{A}_i)-\delta(X_\xi\cdot\mathbf{\Psi})+d(\xi\cdot\mathbf{\Psi}).\nonumber\\
    \label{k_xi}
\end{eqnarray}
Substituting Eq. \eqref{funda2} into Eq. \eqref{k_xi}, we eventually arrive at the following result
\begin{eqnarray}
    d\mathbf{k}_\xi=&-& \delta\alpha^i(\xi\cdot d\mathbf{A}_i)-X_\xi\cdot\boldsymbol{\omega}+\xi\cdot\boldsymbol{\epsilon}{D}^{ab} \delta g_{ab}\nonumber\\
       &-&\xi\cdot\boldsymbol{\epsilon}{E}^{ab} \delta g_{ab}-\xi\cdot\mathbf{E^\prime}_i\delta\alpha^i\nonumber-(\xi\cdot \mathbf{E}'^i)\boldsymbol{\Theta}_i\\
       &-&\delta\mathbf{C}_\xi +\xi\cdot d\mathbf{\Psi}
       +\delta\alpha^i\wedge(X_\xi\cdot\boldsymbol{\Theta}_i).
       \label{fundamental3}
\end{eqnarray}
By evaluating this equation on the solution space where we have 
$\mathbf{E}'^i=0$, $D^{ab}=0$, $d\mathbf{\Psi}=0$, $E^{ab}=0$, and $\mathbf{E}'_i=0$, choosing $\xi$ to be a Killing field but not required to satisfy $\mathcal{L}_\xi\alpha^i=\mathcal{L}_\xi\mathbf{A}_i=0$, and requiring the variations satisfy the linearized equations of motion in the modified theory such that $\delta\mathbf{C}_\xi=0$, we obtain\footnote{Note that $\boldsymbol{\omega}$ and $\mathbf{\Theta}_i$  only involve $\delta g_{ab}$, so we have $X_\xi\cdot \mathbf{\omega}=0$ and $X_\xi\cdot\mathbf{\Theta}_i=0$ due to the fact that $\mathcal{L}_\xi g_{ab}=0$.}
\begin{equation}
    d\mathbf{k}_\xi=- \delta\alpha^i(\xi\cdot d\mathbf{A}_i)=-\delta\alpha^i(\xi\cdot \mathbf{L}_i).
    \label{exfund}
\end{equation}

With the above preparation and by following exactly the same strategy presented in the previous section, we obtain 
\begin{equation}
    \delta [H_\xi]=T\delta S- \delta\alpha^i[\int_{\Sigma}(\xi\cdot \mathbf{L}_i)]+[\int_\mathcal{\infty}\xi\cdot\mathbf{F}].
    \label{ex1stlaw}
\end{equation}
It is noteworthy that the last term of the right-hand side, as a one-form on the solution space of the extended black hole thermodynamics where $\alpha^i$ can be varied, does not vanish generically. But 
as pointed out at the end of the previous section, it turns out to be vanishing when restricted onto the solution space of the ordinary black hole thermodynamics with $\alpha^i$ fixed. Mathematically, this amounts to saying that 
it vanishes on all tangent directions with $\delta \alpha^i=0$,  which implies that it can be expressed in terms of the following one-form 
\begin{equation}
  [\int_\infty\xi\cdot \mathbf{F}]=\mathcal{V}^{(2)}_i\delta\alpha^i
\end{equation}
on the solution space of the extended black hole thermodynamics with $\mathcal{V}^{(2)}_i$ as the corresponding coefficients. 
 So if we further introduce
\begin{equation}
   \mathcal{V}^{(1)}_i\equiv-[\int_\Sigma\xi\cdot\mathbf{L}_i],
\end{equation}
and define $\mathcal{V}_i=\mathcal{V}_i^{(1)}+\mathcal{V}_i^{(2)}$, then we end up with the standard expression for the first law of the extended black hole thermodynamics 
\begin{equation}
    \delta [H_\xi]=T\delta S+ \mathcal{V}_i\delta\alpha^i,
    \label{ex1stlaw2}
\end{equation}
which is exactly the same as that obtained in \cite{Xiao:2025gle}, where the variation of $\alpha^i$ is made by brute force rather than by introducing the auxiliary gauge fields as achieved here.

Note that although we resort to the auxiliary gauge fields $\mathbf{A}_i$ to make the variation of $\alpha^i$ legitimate in the framework of the modified theory, the final expression for the volume terms has no bearing on the auxiliary gauge fields. But in what follows, we shall show that one can also express the volume terms in the form of the auxiliary gauge fields as Eq. \eqref{chemi} proposed in \cite{Hajian:2023bhq}. To this end, we like to choose a  coordinate system $\{x^\mu\}$ such that $\xi=\frac{\partial}{\partial x^0}$ and $x^1$ represents the radial coordinate. Accordingly, we have 
\begin{equation}
    \mathbf{L}_i=L_i(x^1,x^2,\cdots,x^n)\boldsymbol{\varepsilon}
\end{equation}
with $\boldsymbol{\varepsilon}\equiv dx^0\wedge dx^1\wedge\dots\wedge dx^n$, where we have used the fact $\mathcal{L}_\xi\mathbf{L}_i=0$. Motivated by the fact that $d\mathcal{L}_\xi\mathbf{A}_i=\mathcal{L}_\xi\mathbf{L}_i=0$, we like to simplify our life by working with the gauge in which $\mathcal{L}_\xi \mathbf{A}_i=0$ in both the black hole background and the reference spacetime. Such a gauge can always be chosen for the spacetime dimension $n$ greater than one. To demonstrate this, we now present a specific construction for such auxiliary gauge fields as follows
\begin{equation}
    \mathbf{A}_i=A_i^\mu(x^1,x^2,\cdots,x^n)\boldsymbol{\varepsilon}_\mu,
\end{equation}
where $\boldsymbol{\varepsilon}_\mu\equiv(-1)^\mu dx^0\wedge \dots\wedge dx^{\mu-1}\wedge  dx^{\mu+1}\wedge\dots\wedge dx^n$ and $A_i^\mu=A_i(x^1,x^2,\cdots,x^n)\delta_1^\mu$ with $A_i$ determined by $\partial_1A_i=L_i$ up to the residual gauge transformation $A_i\rightarrow A_i+a_i(x^2,x^3,\cdots,x^n)$. 
Under this gauge, we have 
\begin{eqnarray}
    \mathcal{V}_i&=&\mathcal{V}_i^{(2)}+[\int_{\Sigma}d(\xi\cdot \mathbf{A}_i)]\nonumber\\
    &=&\mathcal{V}_i^{(2)}+\int_{\Sigma}d(\xi\cdot \mathbf{A}_i)-\int_{\Sigma^0}d(\xi\cdot A^0_i)\nonumber\\
    &=&\mathcal{V}_i^{(2)}+ [\int_{\infty}(\xi\cdot \mathbf{A}_i) ]-\int_{\mathcal{B}}(\xi\cdot \mathbf{A}_i)\label{chemif}.
\end{eqnarray}
To reach Eq. \eqref{chemi},  one is required to have the first two terms in the last line cancel out, which can be achieved by making use of the above residual gauge choice of $a_i$. The interested readers are referred respectively to Appendix \ref{A} and Appendix \ref{B} for the global conditions and an explicit example for such a gauge choice. 
Thus we have succeeded in validating the proposal for the volume terms made in \cite{Hajian:2023bhq}, although such a proposal requires one to choose a specific gauge as it should be the case.

\section{Conclusion} \label{sec:conclusion}
By employing the Iyer-Wald formalism for the modified theory, where the coupling constants are promoted to the dynamical scalar field by introducing the auxiliary gauge fields, we have accomplished the rigorous derivation for the first law of the extended black hole thermodynamics with the auxiliary gauge field independent expression for the volume terms exactly the same as that obtained in \cite{Xiao:2025gle}, where the coupling constants are varied by brute force instead. In addition, by appropriately choosing a gauge for the auxiliary gauge fields, we have also derived the equivalent expression for the volume terms evaluated on the black hole horizon with the auxiliary gauge field involved, substantiating the proposal made in \cite{Hajian:2023bhq}. 

We conclude our paper by pointing out that although we focus only on the pure gravity in this paper,  the strategy developed here is expected to be also applicable to the gravitational system coupled with matters such as the Maxwell field.

\section*{Acknowledgments}
This work is partially supported by the National Natural Science Foundation of China with Grant Nos. 12361141825,  12375058, and 12575047. 

\newpage
\appendix

\section{Sufficient global conditions for the gauge choice}
\label{A}
Suppose that there exists a global function $t$ such that the $n$-dimensional spacetime region $M$ outside the stationary black hole event horizon can be foliated by the surfaces of constant $t$ and the Killing vector field $\xi$ can be parametrized as $\frac{\partial}{\partial t}$.  Let $\Sigma$ denote the surface of $t=0$ and $\mathcal{C}$ denote the spatial boundary of $\Sigma$ near the infinity. 

The existence of the gauge choice in Eq. \eqref{chemif} amounts to requiring (AI) There exists an $(n-1)$-form $\mathbf{A}_i$ such that $d\mathbf{A}_i=\mathbf{L}_i$ and $\mathcal{L}_\xi \mathbf{A}_i=0$ and 
(AII) There exists an $(n-2)$-form $\boldsymbol{a}$ such that $\mathcal{L}_\xi d\boldsymbol{a}=0$ and $\int_\mathcal{C}\xi\cdot d\boldsymbol{a}\neq0$.
The sufficient conditions can be respectively specified as follows (BI)$\;H^{n-1}(\Sigma)=0$ and (BII)$\;[\mathcal{C}]\in H_{n-2}(\Sigma)$, $[\mathcal{C}]\neq0$,
where $H^p(\Sigma)$ and $H_p(\Sigma)$ denote the $p$-th de Rham cohomology group and the $p$-th singular homology group of $\Sigma$ respectively, and $[\mathcal{C}]$ is the fundamental class of $\mathcal{C}$.

We first prove that (BI) is sufficient for (AI).
As such, note that the homotopy invariance implies that $H^n(M)\cong H^n(R\times \Sigma)\cong H^n(\Sigma)=0$, which tells us the $n$-form $\mathbf{L}_i$ is always exact, namely $\mathbf{L}_i=d\mathbf{B}_i$. In addition, we have
\begin{equation}
  0=\mathcal{L}_\xi\mathbf{L}_i=\mathcal{L}_\xi d\mathbf{B}_i=d\mathcal{L}_\xi\mathbf{B}_i,
\end{equation}
which means that $\mathcal{L}_\xi\mathbf{B}_i$ is a closed $(n-1)$-form. Since
$H^{n-1}(M)\cong H^{n-1}(\mathbb{R}\times\Sigma)\cong H^{n-1}(\Sigma)=0$, it is also exact and can be written as
\begin{equation}\label{eq:I-beta}
  \mathcal{L}_\xi\mathbf{B}_i=d\boldsymbol{\beta}_i.
\end{equation}
Let $\mathbf{A}_i=\mathbf{B}_i+d\boldsymbol{\gamma}_i$, then we have
\begin{equation}\label{eq:I-LieA}
  \mathcal{L}_\xi\mathbf{A}_i
  =\mathcal{L}_\xi\mathbf{B}_i+\mathcal{L}_\xi d\boldsymbol{\gamma}_i
  =d\bigl(\boldsymbol{\beta}_i+\mathcal{L}_\xi\boldsymbol{\gamma}_i\bigr).
\end{equation}
Thus
\begin{equation}\label{eq:I-target}
  \mathcal{L}_\xi\boldsymbol{\gamma}_i=-\boldsymbol{\beta}_i
\end{equation}
can give rise to  $\mathcal{L}_\xi\mathbf{A}_i=0$, which is what we go for.

Eq. \eqref{eq:I-target} is always solvable for $\boldsymbol{\gamma}_i$ on $M$.
Indeed, fix $p\in\Sigma$ and choose local coordinates $\{x^i\}$ on $\Sigma$
in the neighborhood of $p$, then $\{t,x^i\}$ forms a coordinate chart on $M$ covering the entire
integral curve of $\xi$ through $p$. In this chart, Eq. \eqref{eq:I-target}
becomes a set of
first-order ordinary differential equations (ODEs) along $t$ for each coordinate component of $\boldsymbol{\gamma}_i$. Hence a unique
solution exists along the integral curve for any initial data on $p$. By choosing smooth initial data $\boldsymbol{\gamma}_i$ on $\Sigma$ such as 
$\boldsymbol{\gamma}_i|_\Sigma=0$ and solving the ODEs along every integral curve, we further obtain a globally defined smooth $(n-2)$-form $\boldsymbol{\gamma}_i$ on $M$ satisfying
Eq. \eqref{eq:I-target}. \hfill$\square$

Next we prove that (BII) is sufficient for (AII). Since $[\mathcal{C}]\neq0$ in $H_{n-2}(\Sigma)$, de Rham's theorem guarantees the
existence of a closed (${n-2}$)-form $\boldsymbol{\Omega}$ on $\Sigma$ ($d_\Sigma\boldsymbol{\Omega}=0$)
such that
\begin{equation}\label{eq:nonvanishing}
  \int_{\mathcal{C}}\boldsymbol{\Omega} \neq 0 .
\end{equation}
Now let $\pi:M\to\Sigma$ be the canonical projection and  define an (${n-2}$)-form $\boldsymbol{a}$ on $M$ as
\begin{equation}\label{eq:eta-def}
  \boldsymbol{a} = t\pi^*\boldsymbol{\Omega}.
\end{equation}
Whence we have
\begin{equation}\label{eq:deta}
  d\boldsymbol{a}
  = d(t\pi^*\boldsymbol{\Omega})
  = dt\wedge\pi^*\boldsymbol{\Omega} + t d(\pi^*\boldsymbol{\Omega})
  = dt\wedge\pi^*\boldsymbol{\Omega},
\end{equation}
where we have used the fact that  $d(\pi^*\boldsymbol{\Omega})=\pi^*d_\Sigma \boldsymbol{\Omega}=0$. Furthermore, note that $\xi\cdot\pi^*\Omega=0$, so we have
\begin{eqnarray}
   &&\xi\cdot d\boldsymbol{a}
  = (\xi\cdot dt)\wedge\pi^*\boldsymbol{\Omega} - dt\wedge(\xi\cdot\pi^*\boldsymbol{\Omega})
  = \pi^*\boldsymbol{\Omega},\nonumber\\
     &&\mathcal{L}_\xi d\boldsymbol{a}=d\mathcal{L}_\xi\boldsymbol{a}
  = d(\xi\cdot d\boldsymbol{a})
  =d(\pi^*\boldsymbol{\Omega})=0.
\end{eqnarray}
Restricting to $\mathcal{C}\subset\Sigma$ and using Eq. \eqref{eq:nonvanishing}:
\begin{equation}\label{eq:integral}
  \int_{\mathcal{C}}\xi\cdot d\boldsymbol{a}
  = \int_{\mathcal{C}}\pi^*\boldsymbol{\Omega}
  = \int_{\mathcal{C}}\boldsymbol{\Omega} \neq 0 .
\end{equation}
Thus the (${n-2}$)-form $\boldsymbol{a}$ satisfies both
requirements of (AII). \hfill$\square$

As a summary, for any stationary black hole solution satisfying the global conditions (BI) and (BII), there exists a gauge choice that casts the thermodynamic volume into the form of Eq. \eqref{chemi}. 

Actually such conditions are satisfied for a large family of stationary black holes in both asymptotically flat and asymptotically AdS spacetimes, where $\Sigma= \mathbb{R}\times \mathcal{C}$ with $\mathbb{R}$ denoting the radial direction and $\mathcal{C}$ being a compact, connected, and orientable manifold without boundaries. Speaking specifically, we have $H^{n-1}(\mathbb{R}\times \mathcal{C})\cong H^{n-1}(\mathcal{C})=0$, which means that (BI) is satisfied. On the other hand, we also have $H_{n-2}(\mathbb{R}\times \mathcal{C})\cong H_{n-2}(\mathcal{C})$. Note that for any compact, connected, and orientable manifold without boundaries, the highest order singular homology group is the group generated by its fundamental class. Thus $H_{n-2}(\mathcal{C})\cong \mathbb{R}$ and is generated by $[\mathcal{C}]\neq0$ in $H_{n-2}(\mathcal{C})$, which further implies $[\mathcal{C}]\neq0$ in $H_{n-2}(\mathbb{R}\times\mathcal{C})$. Therefore, (BII) is also satisfied.

\section{An explicit example for the gauge choice}\label{B}
We take as an example the rotating Banados-Teitelboim-Zanelli (BTZ) black hole solution
\begin{equation}
    ds^2=-\Delta dt^2+\frac{dr^2}{\Delta}+r^2(d\phi-\frac{j}{2r^2}dt)^2\label{BTZmetric}
\end{equation} 
in the new massive gravity theory with the Lagrangian form given by 
\begin{equation}
    \mathbf{L}=\frac{1}{16\pi}\bigg[ R-2\Lambda-\beta\bigg( \frac{3}{8}R^2-R_{\mu\nu}R^{\mu\nu}\bigg)\bigg]\boldsymbol{\varepsilon},
\end{equation}
where $\Delta=-m+\frac{r^2}{l^2}+\frac{j^2}{4r^2}$ and the AdS radius $l$ is determined by $\Lambda=-\frac{1}{l^2}+\frac{\beta}{4l^2}$. The event horizon satisfies $2r_h^2=l^2(m+\sqrt{m^2-\frac{j^2}{l^2}})$ with the Killing vector field chosen to be $\xi=\frac{\partial}{\partial t}+\Omega_H\frac{\partial}{\partial \phi}$ normal to the horizon. We choose the reference spacetime as
\begin{equation}
    ds^2=-\Delta_0\lambda^2(r_c) dt^2+\frac{dr^2}{\Delta_0}+r^2d\phi,
\end{equation}
where $\Delta_0=\frac{r^2}{l^2}$ and $\lambda(r_c)=\sqrt\frac{\Delta(r_c)}{\Delta_0(r_c)}$ with $r_c$ the cutoff radius near the infinity. Let $\alpha^\Lambda=-\frac{\Lambda}{8\pi}$ and $\alpha^\beta=-\frac{\beta}{16\pi}$, then the modified Lagrangian form can be written as
\begin{equation}
     \mathbf{L}'= \mathbf{L}_0+\alpha^\Lambda(x) (\mathbf{L}_\Lambda-d\mathbf{A}_\Lambda)+\alpha^\beta(x) (\mathbf{L}_\beta-d\mathbf{A}_\beta)
\end{equation}
with $\mathbf{L}_\Lambda=\boldsymbol{\varepsilon}$ and $\mathbf{L}_\beta=\bigg( \frac{3}{8}R^2-R_{\mu\nu}R^{\mu\nu}\bigg)\boldsymbol{\varepsilon}$. The on-shell Lagrangian forms can be calculated out as 
\begin{eqnarray}
    \mathbf{L}_\Lambda=rdt\wedge dr\wedge d\phi,\quad
    \mathbf{L}_\beta=\frac{3r}{2 l^4}dt\wedge dr\wedge d\phi
\end{eqnarray}
for the BTZ black hole solution \eqref{BTZmetric}. 
The on-shell Lagrangian forms of the reference metric are given simply by $\mathbf{L}_\Lambda^0=\lambda(r_c)\mathbf{L}_\Lambda$, $\mathbf{L}_\beta^0=\lambda(r_c)\mathbf{L}_\beta$. Accordingly, we have
\begin{eqnarray}
    \mathcal{V}^{(1)}_\Lambda&&\equiv-[\int_\Sigma\xi\cdot\mathbf{L}_\Lambda]=\pi r_h^2-\frac{\pi l^2m}{2},\nonumber\\
    \mathcal{V}^{(1)}_\beta&&\equiv-[\int_\Sigma\xi\cdot\mathbf{L}_\beta]=\frac{3\pi}{2l^4}( r_h^2-\frac{ l^2m}{2}).\label{BTZv1}
\end{eqnarray}
Using the general definition of $\mathbf{F}$ for the $f(R_{abcd})$ theory \cite{Guo:2025ohn}, one can obtain \cite{Xiao:2025gle} 
\begin{eqnarray}
     [\int_\infty\xi\cdot \mathbf{F}]&=&\frac{\pi l^2m(2l^2+\beta)}{2(2l^2-\beta)}\delta\alpha^\Lambda-\frac{m\pi(2l^2+\beta)}{4l^2(2l^2-\beta)}\delta\alpha^\beta\nonumber\\
&=&\mathcal{V}^{(2)}_\Lambda\delta\alpha^\Lambda+\mathcal{V}^{(2)}_\beta\delta\alpha^\beta,
\end{eqnarray}
whereby one can read off
\begin{equation}
    \mathcal{V}_\Lambda^{(2)}=\frac{\pi l^2m(2l^2+\beta)}{2(2l^2-\beta)}, \;\;\;\;\;
    \mathcal{V}_\beta^{(2)}=-\frac{m\pi(2l^2+\beta)}{4l^2(2l^2-\beta)}.\label{BTZF}
\end{equation}
Combining Eq. \eqref{BTZF} and Eq. \eqref{BTZv1} yields the following thermodynamic volumes
\begin{eqnarray}
     \mathcal{V}_\Lambda=\pi r_h^2+\frac{\pi\beta l^2m}{2l^2-\beta},\;\;\;\;\;
      \mathcal{V}_\beta=\frac{3\pi r_h^2}{2l^4}-\frac{\pi m(4l^2-\beta)}{2l^2(2l^2-\beta)}.\nonumber\\
\end{eqnarray}

To cast the thermodynamic volumes into the form of Eq. \eqref{chemi}, we can first choose the auxiliary gauge fields as follows
\begin{align}
    \mathbf{A}_\Lambda=-\frac{r^2}{2}dt\wedge d\phi+d\mathbf{a}_\Lambda,\quad
    \mathbf{A}_\beta=-\frac{3r^2}{4l^4}dt\wedge d\phi+d\mathbf{a}_\beta,\nonumber
    \\
\end{align}
up to the residual gauge transformations given by $\mathbf{a}_\Lambda$ and $\mathbf{a}_\beta$ with $\mathcal{L}_\xi d\mathbf{a}_\Lambda=\mathcal{L}_\xi d\mathbf{a}_\beta=0$. For the reference spacetime, we simply fix $\mathbf{A}_\Lambda^0=-\frac{r^2}{2}\lambda(r_c) dt\wedge d\phi$ and $\mathbf{A}_\beta^0=-\frac{3r^2}{4 l^4}\lambda(r_c) dt\wedge d\phi$. 
We can further compute
\begin{equation}
    \mathcal{V}_\Lambda^{(2)}+ [\int_{\infty}(\xi\cdot \mathbf{A}_\Lambda)]=\int_{\infty}(\xi\cdot d\mathbf{a}_\Lambda)+\frac{\pi ml^2\beta}{2l^2-\beta}.\label{alambda}
\end{equation}
If there is a one-form $\mathbf{a}$ such that $\mathcal{L}_\xi d\mathbf{a}=0$ and $\int_{\infty}\xi\cdot d\mathbf{a}\neq 0$. One can always choose $\mathbf{a}_\Lambda=c_\Lambda\mathbf{a}$ for some constant $c_\Lambda$ to make Eq. \eqref{alambda} vanish. Such an $\mathbf{a}$ is highly non-unique, and the simplest one can be
\begin{equation}
    \mathbf{a}=td\phi,
\end{equation}
which gives rise to $\int_{\infty}\xi\cdot d\mathbf{a}=2\pi$. To make Eq. \eqref{alambda} vanish, we choose
\begin{equation}
    c_\Lambda=\frac{m\beta l^2}{2\beta-4l^2}.
\end{equation}
$\mathbf{a}_\beta$ can be chosen in the same way. The final results are given by 
\begin{equation}
    \mathbf{a}_\Lambda=\frac{m\beta l^2}{2\beta-4l^2}td\phi,\;\;\;\;\;\;\;\mathbf{a}_\beta=\frac{m(\beta-4l^2)}{4l^2(\beta-2l^2)}td\phi.
\end{equation}
These are exactly the same gauge choices as in \cite{Hajian:2023bhq}, where the thermodynamic volumes can be expressed in the form of Eq. \eqref{chemi}.

\vfill

\end{CJK*}

\end{document}